\documentclass[journal,comsoc]{IEEEtran}

\usepackage[english]{babel}
\usepackage[utf8]{inputenc}
\usepackage{amsmath}
\usepackage{amssymb}
\usepackage{graphicx}
\usepackage{booktabs}
\usepackage{multirow}
\usepackage{color}
\usepackage{url}
\usepackage[absolute]{textpos}
\makeatletter
\let\MYcaption\@makecaption
\makeatother
\usepackage[font=footnotesize]{subcaption}
\makeatletter
\let\@makecaption\MYcaption
\makeatother

\newcommand{\AZ}[1]{}

\title{Uncoordinated access schemes for the IoT: approaches, regulations, and performance}
\author{Daniel Zucchetto, Andrea Zanella \\
Department of Information Engineering, University of Padova, Italy \\
E-mail: \{\emph{zucchett,zanella}\}@dei.unipd.it}

\begin{document}
\maketitle

\begin{abstract}
Internet of Things~(IoT) devices communicate using a variety of protocols, differing in many aspects, with the channel access method being one of the most important. Most of the transmission technologies explicitly designed for IoT and Machine-to-Machine~(M2M) communication use either an ALOHA-based channel access or some type of Listen Before Talk (LBT) strategy, based on carrier sensing. In this paper, we provide a comparative overview of the uncoordinated channel access methods for IoT technologies, namely ALOHA-based and LBT schemes, in relation with the ETSI and FCC regulatory frameworks. Furthermore, we provide a performance comparison of these access schemes, both in terms of successful transmissions and energy efficiency, in a typical IoT deployment. Results show that LBT is effective in reducing inter-node interference even for long-range transmissions, though the energy efficiency can be lower than that provided by ALOHA methods. The adoption of rate-adaptation schemes, furthermore, lowers the energy consumption while improving the fairness among nodes at different distances from the receiver.
Coexistence issues are also investigated, showing that in massive deployments LBT is severely affected by the presence of ALOHA devices in the same area.
\end{abstract}

\begin{textblock*}{17cm}(1.7cm,0.5cm)
\noindent\scriptsize The final version of this paper has been published in the IEEE Communications Magazine vol. 55, no. 9, pp. 48--54, September 2017. DOI: 10.1109/MCOM.2017.1600617\\
\textbf{Copyright Notice}: \textcopyright 2017 IEEE. Personal use of this material is permitted. Permission from IEEE must be obtained for all other uses, including reprinting/republishing this material for advertising or promotional purposes, collecting new collected works for resale or redistribution to servers or lists, or reuse of any copyrighted component of this work in other works.
\end{textblock*}

\section{Introduction}

A key element to enable the full realization of the Internet of Things (IoT) vision is the ubiquitous connectivity of end devices, with minimal configuration, as for the so-called \emph{place-\&-play} paradigm \cite{Biral20151}. Today, the main three approaches to provide connectivity to the IoT devices are the following.

\textit{Cellular systems.} The existing cellular networks are a natural and appealing solution to provide connectivity to IoT end-devices, thanks to their world-wide established footprint and the capillary market penetration. Unfortunately, current cellular network technologies have been designed targeting wideband services, characterized by few connections  that generate a large amount of data, while most IoT services are expected to generate a relatively small amount of traffic, but from a very large number of different devices. This shift of paradigm challenges the control plan of current cellular standards, which can become the system bottleneck. For these reasons, the IoT and Machine-to-Machine (M2M) scenarios are considered as major challenges for next generation wireless cellular systems, commonly referred to as 5G.

\textit{Short-range multi-hop technologies.} This family collects a number of popular technologies specifically designed for M2M communications or Wireless Personal Area Networks (WPANs). These systems usually operate in the frequency bands centered around 2.4 GHz, 915 MHz and 868 MHz, though the 2.4 GHz is the most common choice. They are characterized by high energy efficiency and medium/high bitrates (order of hundreds of kbit/s or higher), but limited single-hop coverage area. To cover larger areas, most WPAN technologies provide the possibility to relay data in a multihop fashion, realizing a so-called \emph{mesh network}. Examples of standards in this category are IEEE~802.15.4~\cite{15.4}, Bluetooth Low Energy~\cite{BLE}, and Z-Wave, the latter having its physical and data link layers specified in ITU-T~G.9959~\cite{g.9959}.

\textit{Low-Power Wide-Area (LPWA) networks.} A third relevant class in the arena of IoT-enabling wireless technologies consists in the LPWA solutions. According to~\cite{cisco}, LPWA technologies will account for 28\% of M2M connections by 2020. These technologies, specifically designed to support M2M connectivity, provide low bitrates, low energy consumption, and wide geographical coverage. Almost all  LPWA technologies operate at frequencies around 800 or 900 MHz, though there are also solutions working in the classic 2.4 GHz ISM band or exploiting white spaces in TV frequencies. Some relevant LPWA technologies are LoRaWAN\texttrademark, Sigfox, Ingenu~\cite{CVZZ16}.

While cellular systems entail centralized access schemes over dedicated frequency bands, which provide high efficiency, robustness, security, and performance predictability, most of WPAN and LPWA technologies operate on unlicensed radio bands,
adopting uncoordinated access schemes. The use of unlicensed bands yields the obvious advantage of lowering the operational costs of the network, while the adoption of uncoordinated channel access schemes makes it possible to simplify the hardware of the nodes, thus reducing the manufacturing costs and the energy consumption.
The downside is that the lack of coordination in channel access may yield performance losses in terms of throughput and energy efficiency when the number of contending  nodes increases.  

To alleviate the problem of channel congestion in the unlicensed bands, radio spectrum regulators have imposed limits on the channel occupation of each device, in terms of bandwidth, time, and on the maximum transmission power. However, the Federal Communications Commission (FCC) in the USA and the Conference of Postal and Telecommunications Administrations (CEPT) in Europe have taken different approaches to limit channel congestion: the first imposes very strict limits on the emission power and favors the use of spread spectrum techniques but do not restrict the number of access attempts that can be performed by the nodes \cite{fcc-part15}, while the second limits the fraction of on-air time of a device to be lower than a given \emph{duty cycle}, or imposes the use of \emph{Listen Before Talk} (LBT) techniques, which are also referred to as  \emph{Carrier Sense Multiple Access} (CSMA) protocols \cite{etsi-2012}.\footnote{The two terms will be used interchangeably in this paper.}

These precautions are actually effective when the coverage range of the wireless transmitters is relatively small (few meters), as was indeed the case for the first commercial products operating in the ISM frequency bands. However, this condition does no longer hold for LPWA solutions, which have coverage ranges in the order of 10--15 km in rural areas, and 2--5 km in urban areas, with a star-like topology that can exacerbate the mutual interference and hidden node problems. Furthermore, while short-range communication systems usually support a single, or just a few modulation schemes and transmit rates, LPWA technologies usually provide multiple transmit rates to optimize the transmission based on the distance to be covered. 

Despite these quite radical changes in the transmit characteristics of the recent LPWA technologies with respect to the previous generation of  the so-called Short Range Devices~(SRD), the channel access methods and the regulatory constraints are still the same. The objective of this study is hence to investigate the performance of well established uncoordinated channel access schemes in this new scenario, characterized by a huge number of devices with large coverage ranges and multi-rate capabilities.
To this end, we first provide a quick overview of the main uncoordinated access schemes used by most common wireless communication technologies considered for the IoT and we discuss the regulatory frameworks, with particular focus to the European case. We then compare the performance achieved by two popular uncoordinated access schemes in a typical LPWA network scenario, considering the limits imposed by the regulations.
The paper is then closed with some final considerations and recommendations. 

\section{Uncoordinated access techniques for the IoT}\label{unc}
Channel access schemes can be roughly divided in two main categories: coordinated and uncoordinated (or contention-based). Coordinated access schemes require time synchronization among the nodes and, hence, are more suitable for small networks (e.g., Bluetooth) or centrally controlled systems (e.g., cellular), with predictable and/or steady traffic flows (e.g., voice or bulk data transfer). Uncoordinated access strategies, instead, are usually considered for networks with a variable number of devices and unpredictable traffic patterns. In the following we provide a quick overview of the two main uncoordinated access schemes that are widely adopted by the transmission technologies typically associated to the IoT scenarios. 

\subsection{ALOHA-based schemes}
Many protocols for M2M communication are based on pure ALOHA access schemes, according to which a transmission is attempted whenever a new message is generated by the device. This form of channel access may be coupled with a retransmission scheme, according to which a packet is retransmitted until acknowledged by the receiver. However, some IoT services (e.g., environmental monitoring) can tolerate a certain amount of lost messages. In these cases, a retransmission scheme is not needed, allowing for a simplification of the device firmware and enabling a significant reduction in the energy consumption. For these reasons, ALOHA schemes are widely adopted in M2M communication as, for example, LoRaWAN and Sigfox. Furthermore, some standards that adopt LBT access techniques optionally provide an ALOHA mode of operation, as for the IEEE~802.15.4.

More sophisticated ALOHA-based protocols can be enabled when nodes are time synchronized, e.g., by means of beacons periodically broadcasted by coordinator nodes (e.g., gateways in LoRaWAN). For example, slotted-ALOHA divides the time in intervals of equal size, called slots, and allows transmissions only within slots, thus avoiding packet losses due to partially overlapping transmissions. Framed slotted ALOHA (FSA), instead, organizes the slots in groups, called frames, and allow each node to transmit only once per frame. The limit of these schemes is that packet transmission time should not exceed the slot duration. A common solution to accommodate uneven packet transmission times is to adopt a hybrid access scheme (HYB) that splits the frame in two parts: the first $k$ slots are used by the nodes to send  resource reservation messages to the controller, using a FSA access scheme, while the remaining slots in the frame are allocated by the controller to the nodes, according to the amount of resources required in the accepted reservation messages. The nodes get notified about the allocated resources by a control message that is broadcasted by the controller right after the end of the reservation phase. Variants of these basic mechanism are currently used in many different protocols as, e.g., GSM, 802.11e. However, to the best of our knowledge, the HYB approach has not yet been studied in the M2M  scenario.

\subsection{Carrier sensing schemes}
When using carrier sensing techniques, each device listens to the channel before transmitting (from which the wording ``Listen-Before-Talk''). The channel sensing operation is typically called \emph{Clear Channel Assessment} (CCA) and aims at checking the occupancy of the channel by other transmitters, in which case the channel access will be delayed to avoid mutual interference that may result in the so-called \emph{packet collisions}. The LBT schemes can differ in the way the CCA is performed and in the adopted behavior in case the channel is sensed busy.

The three most common methods to perform the CCA are the following. 
\begin{itemize}
\item \emph{Energy detection} (ED). The channel is detected as busy if the electromagnetic energy on the channel is above a given ED threshold.
\item \emph{Carrier sense} (CS). The channel is reported as busy if the device detects a signal with modulation and spreading characteristics compatible with those used for transmission, irrespective of the signal energy.
\item \emph{Carrier sense with energy detection} (CS+ED). In this case, a logical combination of the above methods is used, where the logical operator can be AND or OR.
\end{itemize}

The IEEE~802.15.4 standard supports all these CCA methods, along with pure ALOHA and two other modes specific for ultra-wideband communications.
In an unslotted system, the backoff procedure for the IEEE~802.15.4 CCA mechanism tries to adapt to the channel congestion by limiting the rate at which subsequent CCAs are performed for the same message. If the number of consecutive backoffs exceeds a given threshold, the message is discarded. Details about the CCA procedure in IEEE 802.15.4 networks can be found in~\cite{15.4}, together with recommendations about the ED threshold and CCA detection time.

\section{The regulatory framework}\label{reg}
The use of unlicensed frequency bands by radio emitters is subject to regulations that are intended to favor the coexistence of a multitude of heterogeneous radio transceivers in the same frequency bands, limiting the mutual interference and avoiding any monopolization of the spectrum by single devices. The radio emitters operating in the ISM frequency bands are typically referred to as ``Short Range Devices.'' However, the ERC Recommendation 70-03, emanated by the CEPT, specifies that \emph{The term Short Range Device (SRD) is intended to cover the radio transmitters which provide either uni-directional or bi-directional communication which have low capability of causing interference to other radio equipment.} Despite the name, there is no explicit mention of the actual coverage range of such technologies. Therefore, long-range technologies operating in the ISM bands, such as Sigfox or LoRa, are still subject to the same regulatory constraints that apply to the actual short range technologies, as IEEE~802.15.4, Bluetooth, IEEE~802.11, and so on.

In the European Union, the European Commission designated the CEPT to define technical harmonization directives for the use of the radio spectrum.
In 1988, under the patronage of the CEPT, the European Telecommunications Standards Institute (ETSI) was created to develop and maintain Harmonized Standards for telecommunications.

In the unlicensed radio spectrum at 868 MHz, the ETSI mandates a duty cycle limit between 0.1\% and 1\% over a 1 hour interval for devices that do not adopt LBT \cite{etsi-2012}.  Only very specific applications, such as wireless audio, are allowed to ignore the duty cycle limitation. The duty cycle constraint can be relaxed by employing an LBT access scheme together with the Adaptive Frequency Agility (AFA), i.e., the ability to dynamically changing channel \cite{etsi-2012}.
Devices with LBT and AFA capabilities, in fact, are only subject to a 2.8\% duty cycle limitation for any 200~kHz spectrum.
An example of technology that adopts the LBT approach is the IEEE 802.15.4 that, however, does not perfectly match the ETSI specifications, since its channel sensing period is shorter than that mandated by ETSI, which is between $5$~ms and $10$~ms, depending on the used bandwidth \cite{etsi-2012}. Instead, the recommendations on the LBT sensitivity, which shall be between $-102$~dBm and $-82$~dBm, are usually satisfied by commercial transceivers.

Due to the adoption by the European Union of a new set of rules for the radio equipments, called Radio Equipment Directive (RED)~\cite{red}, ETSI is reviewing the related Harmonized Standards.
However, devices that are compliant with the previous Radio and Telecommunication Terminal Equipment (R\&TTE) Directive~\cite{rtte} can be placed on the market until June~17,~2017. Furthermore, devices that do not satisfy the constraints imposed by the Harmonized Standards can still be commercialized, but subject to a more comprehensive certification procedure attesting that the device meets the essential requirements of the European Directives~\cite{red}.
The latest draft version of the ETSI Harmonized Standards~\cite{etsi-2016} includes some changes on the medium access procedures. In particular, the LBT technique is generalized as a \emph{polite spectrum access} technique, while AFA is no more required. Furthermore, the LBT ED threshold has been relaxed, while the minimum CCA listening period has been increased.

The agency designated to regulate radio communications in the USA is the FCC, which
also grants permits for the use of licensed radio spectrum and emanates regulations for wired communications.
The FCC regulation does not impose any duty cycle restrictions to emitters operating in the 902--928 MHz band, but limits the maximum transmit power, for non-frequency hopping systems, to $-1.25$~dBm  \cite{fcc-part15}, which is significantly lower than the 14~dBm allowed by ETSI.

\section{Performance analysis}\label{perf}
ALOHA schemes and channel sensing techniques have been comprehensively modeled and their performance limits in terms of throughput and capacity are well understood (see, e.g., \cite{stochastic,kaynia}, just to cite few). However, the use of different spreading techniques and/or modulation-\&-coding-schemes to cope with the interference and to trade transmission speed for reliability, the large coverage range enabled by the LPWA technologies,
the total reuse of the same frequency bands by different technologies, and the limitations imposed by the regulations to the channel access, raise the question on how effective are the classical uncoordinated channel access techniques to adequately support the expected growth of the IoT services. 

In this section we shed some light on these aspects by presenting a simulation analysis of the performance achieved by  ALOHA-based (specifically, pure ALOHA and HYB) and LBT access schemes in the simplest IoT scenario sketched in Figure~\ref{fig:scenario}: a gateway (GW) receiving packets from a multitude of peripheral devices randomly spread over a wide area. Despite its simplicity, this scenario embodies most of the problems that can be expected in a real IoT deployment based on long-range technologies. In particular, we are interested in investigating how the distance from the gateway may impact the performance experienced by the node, with and without multirate capability and using either ALOHA or LBT techniques. ALOHA-based access schemes, in fact, allow the maximum energy saving in light traffic conditions, since they avoid the (even small) energy cost involved in carrier sensing. On the other hand, nodes farther away from the gateway are likely more prone to transmission failure due to interference, which however can potentially be mitigated by the use of LBT. Furthermore, the adoption of rate adaptation techniques is expected to increase the system capacity by reducing the transmit time of nodes closer to the gateway that not only will experience a lower interference probability, but will also have the chance to transmit more packets within the duty cycle limitations. It is hence interesting to investigate how much of such a performance gain will be transferred to the more peripheral nodes, and whether the LBT techniques can further improve performance in a significant manner.   

\subsection{Simulation scenario}
\begin{table}
	\centering
	\caption{Simulation parameters}
	\label{tab:param}
	\begin{tabular}{llr}
		\toprule
		\multicolumn{2}{c}{Parameter} & \multicolumn{1}{c}{Value} \\
		\midrule
		Spatial node density & $\lambda_s$ & $10^{-3}$ nodes/m$^2$ \\
		Packet generation rate & $\lambda_t$ & 0.01 packets/s \\
		Transmission power & $P_\mathrm{TX}$ & 14~dBm \\
		Transmission frequency & $f$ & 868~MHz \\
		Path loss coefficient & $A$ & 36.36~m$^{-1}$ \\
		Path loss exponent & $\beta$ & 3.5 \\
		Packet length & $L$ & 240~bit \\
		Transmission bitrates & $\mathcal{R}$ & $\{0.5, \ldots,100\}$~kbit/s \\
		Bandwidth & $B_W$ & 400~kHz \\
		Noise spectral density & $N_0$ & $2\cdot10^{-20}$ W/Hz \\
		Duty cycle & $\delta_T$ & 1\% \\
		Circuit power & $P_c$ & 16~dBm \\
		Sensing time & $T_s$ & 0.4 ms \\		
		\multirow{2}{*}{Sensing energy} & \multirow{2}{*}{$E_s$} & 3.98~$\mu$J (LBT) \\
		 & & 0.2~mJ (LBT+ETSI) \\
		Smoothing parameter& $\alpha$ &0.1\\
		Target outage probability for RA & $p^*$ & 0.05\\ 
		\midrule
		\multicolumn{3}{l}{\emph {HYB parameters}} \vspace{0.1em}\\
		Frame duration& $T_W$ & 60~s \\
		Number of reservation slots in a frame & $N_{RM}$ & 80 \\
		Reservation message size & $L_{RM}$ & 24~bits \\
		Reservation message transmit rate & $R_{RM}$ & 500~bit/s \\
		Beacon duration & $T_B$ & 0.12~s \\
		Resource notification message duration& $T_{RA}$ & 3.84~s \\
		\bottomrule
	\end{tabular}
\end{table}

\begin{figure}
	\centering
	\includegraphics[width=0.75\linewidth]{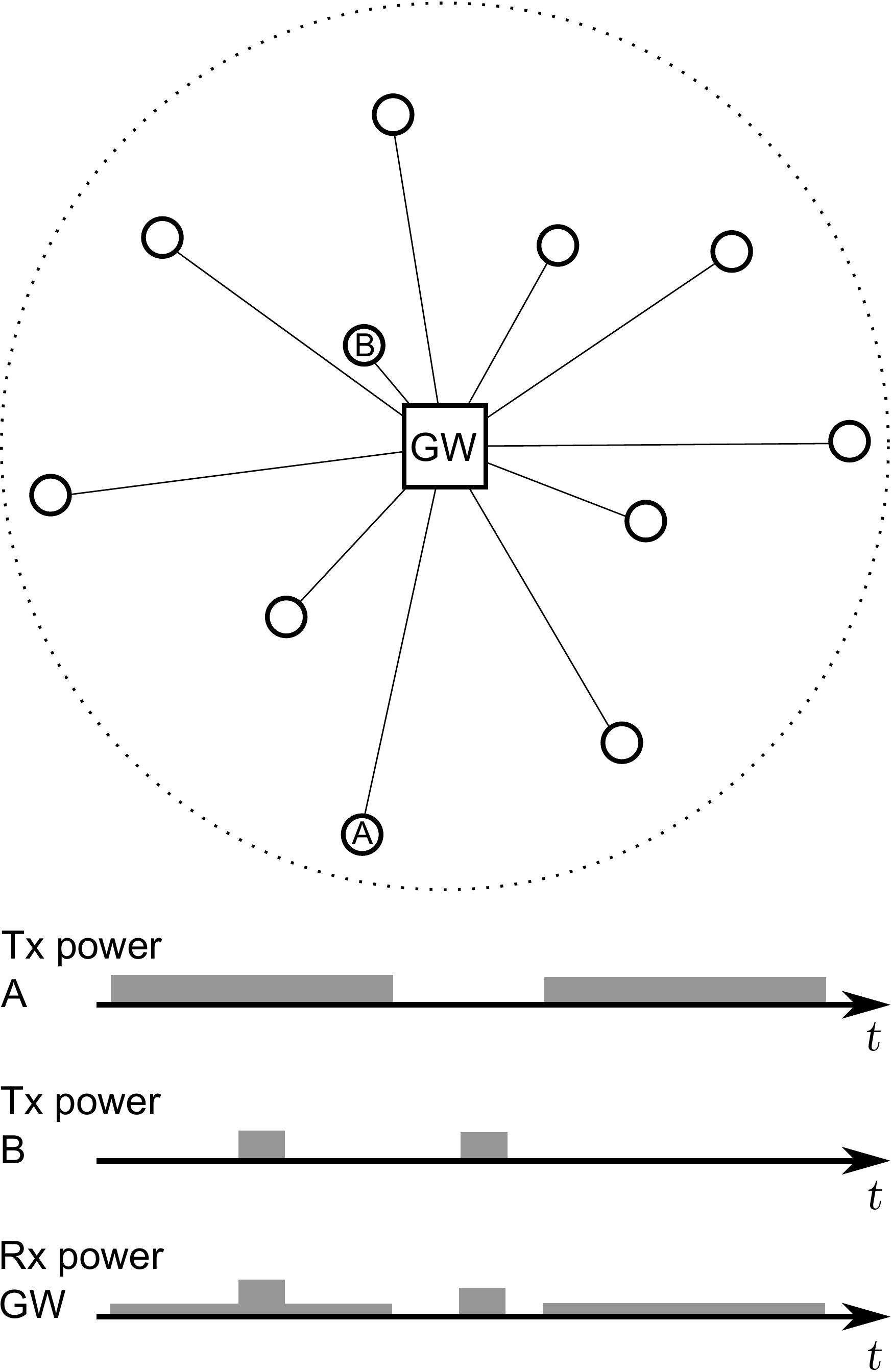}
	\caption{Above: simulation scenario, with multiple  transmitters scattered around the common receiver (GW). Below: example of signal transmissions by nodes A and B, using different bitrates, and of received signal power at the gateway.}
	\label{fig:scenario}
\end{figure}

In our simulations 
we consider a propagation model given by the product of the channel gain, $\gamma(d)=(Ad)^{-\beta}$, which accounts for the power decay with the distance $d$ from the transmitter through the model parameters $A$ and $\beta$, and the Rayleigh fading gain, which is modelled as an exponential random variable with unit mean.

We consider a limited set of possible transmission rates, namely $\mathcal{R}=\{0.5,1,5,10,50,100\}$~kbit/s, and assume that a packet transmitted at rate $r \in \mathcal{R}$ is correctly decoded if the received signal energy over the total noise energy plus interference energy collected by the receiver during the packet reception time (i.e., the Signal-to-Interference-and-Noise Ratio, SINR) is above a certain threshold $\Gamma_{th}(r)$, which is determined from the Shannon channel capacity as
\begin{equation}
\Gamma_{th}(r) = 2^{r/W}-1
\label{th}
\end{equation}
where $W$ is the signal bandwidth.

For the single rate case~(SR), we suppose that all nodes transmit with the lowest bitrate of 500~bit/s. For the multirate scenario, instead, we consider a simple rate-adaptation mechanism that keeps a moving-average estimate of the SINR (using a smoothing factor $\alpha$) and selects the rate $R$ so that the expected outage probability is not larger than $p^*=0.05$. To improve the energy efficiency, furthermore, we assumed no acknowledgement or retransmission mechanism is implemented, so that packets that are not successfully received are definitely lost.

The LBT scheme has been implemented based on the IEEE~802.15.4 specifications. The ED CCA threshold has been chosen to match the minimum signal power required to correctly receive a packet transmitted at the basic rate of $500$~bit/s. This value is compatible with the limits on the LBT threshold imposed by ETSI~\cite{etsi-2012}.

As exemplified in Figure~\ref{fig:scenario}, transmitting nodes are distributed as for a spatial Poisson process of rate~$\lambda_s$ [devices/m$^2$] over a circle with radius equal to the maximum coverage distance at the basic rate of $500$ bit/s. Each device generates messages of length $L$ according to a Poisson process of rate~$\lambda_t$ [packets/s]. All messages are addressed to the gateway that is placed at the center of the circle.

The setting of all the simulation parameters is reported in Table~\ref{tab:param}.

\begin{figure}
	\centering
	\includegraphics[width=1\linewidth]{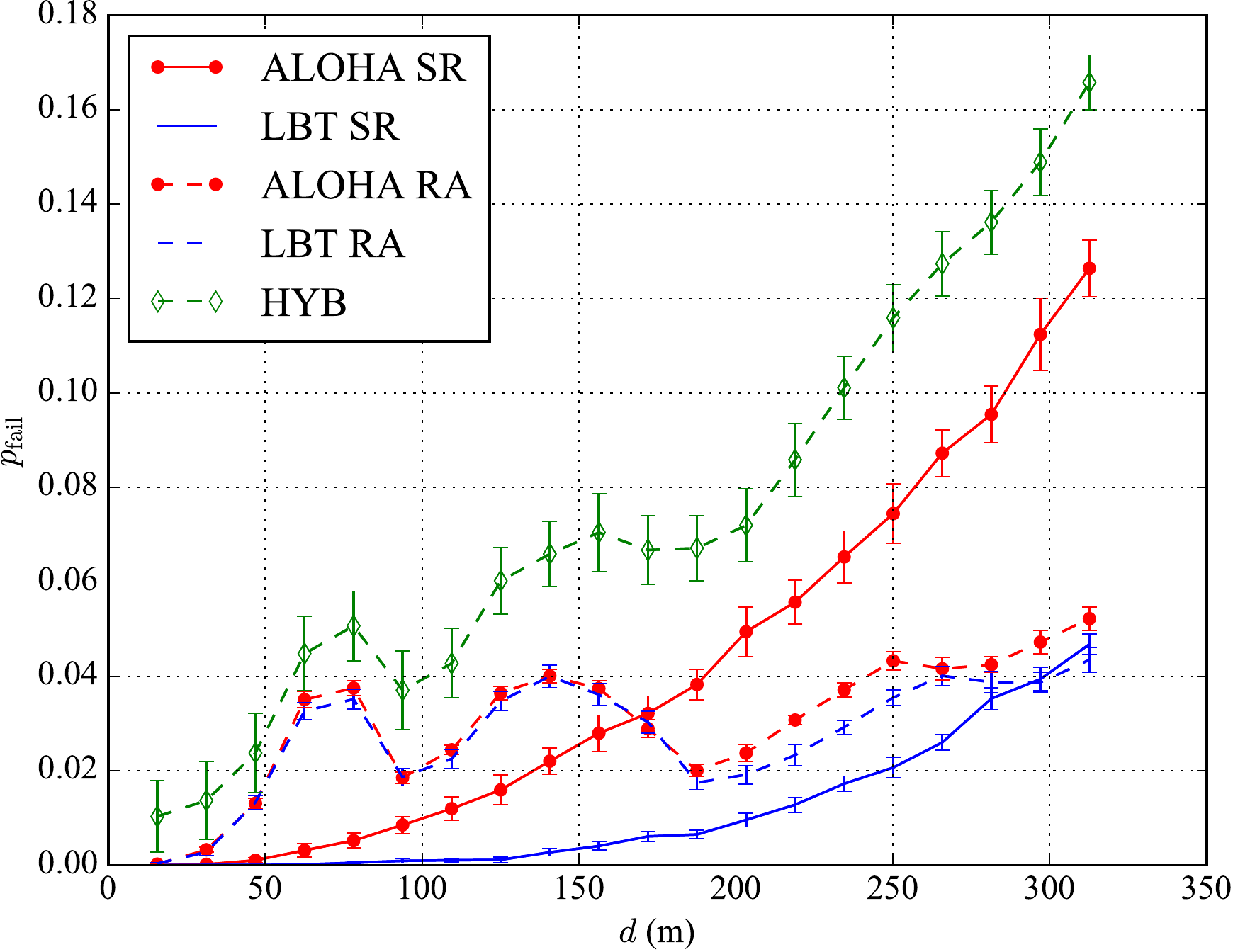}
	\caption{$p_\mathrm{fail}$ for ALOHA and LBT, for single rate (SR) and rate adaptive (RA) cases, with 95\%~confidence intervals.}
	\label{fig:pfail}
\end{figure}

\subsection{Transmission failure probability}
We define  $p_\mathrm{fail}$ as the probability that a transmitted message (including
reservation messages in case of HYB) is received with SINR
below threshold and, hence, is not correctly decoded. For HYB we also
include in the  $p_\mathrm{fail}$ the transmission requests that are not accepted
because of lack of slots in the transmission part of the frame.
Note that, while we consider both the Single rate (SR) and Rate Adaptation (RA) versions of the pure-ALOHA and LBT schemes, for the HYB protocol we only consider the RA version, since this access scheme is more effective when packet transmissions have uneven duration.
In Figure~\ref{fig:pfail} we report the failure probability for target nodes placed at increasing distances from the gateway. Red curves with circle markers refer to ALOHA, blue plain curves to LBT, and green dashed line with diamond markers to HYB. Solid and dashed lines have been associated to the SR and RA case, respectively.

For the SR case, we can see that the failure probability grows with the distance from the gateway, since nodes farther away have less SINR margin for successful decoding and are hence less robust to the interference produced by overlapping transmissions. In this case, carrier sense can indeed improve performance, even if the sensing range does not prevent the hidden node problem.

The downside of using LBT (not reported here for space constraints) is that up to 55\% of the transmission attempts are aborted, in high traffic conditions, because the maximum number of CCAs is reached without finding an idle channel. 

The adoption of RA changes significantly the performance, smoothing out the differences between the two access protocols. Indeed, higher bitrates allow the nodes near the receiver to occupy the channel for a lower period of time, thus reducing the probability of overlapping with other transmissions and improving the performance of both access schemes.
Note that the change of rate with the distance is reflected by the oscillation in the failure probability that, however, remains approximately below $1-p^*$.

Rather interestingly, HYB performs worse than the other schemes. The reason is that, in the considered scenario, the transmit time of reservation messages, always sent at the basic rate, is comparable to that of data packets sent at higher rates. Therefore, the reservation channel can become the system bottleneck. The overall channel occupancy of HYB is thus significantly higher than that of the other two schemes, yielding higher failure probability.

\subsection{Energy efficiency}
Another key performance index in the IoT scenario is the \emph{energy efficiency}, which is here defined as the ratio of the total number of bits successfully delivered to the gateway over the entire energy consumed by the node (including channel sensing and failed transmissions).

We modelled the power consumed during a transmission as the sum of a constant term, named circuit power, that represents the power used by the radio circuitry, and a term that accounts for the radiated power, which is called transmission power. When using LBT, we also add the power required to perform the ED CCA.
Referring to the data-sheets of some off-the-shelf modules,\footnote{Atmel AT86RF212B, Texas Instruments CC1125 and CC1310, and Semtech SX1272 modules.} we set the circuit power to 16~dBm, the transmit power to 14~dBm, the receive power to 13~dBm, and the CCA power to 10~dBm~\cite{power-ed, negri2005flexible}.

\begin{figure}
	\centering
	\begin{subfigure}{1\linewidth}
		\centering
		\includegraphics[width=1\linewidth]{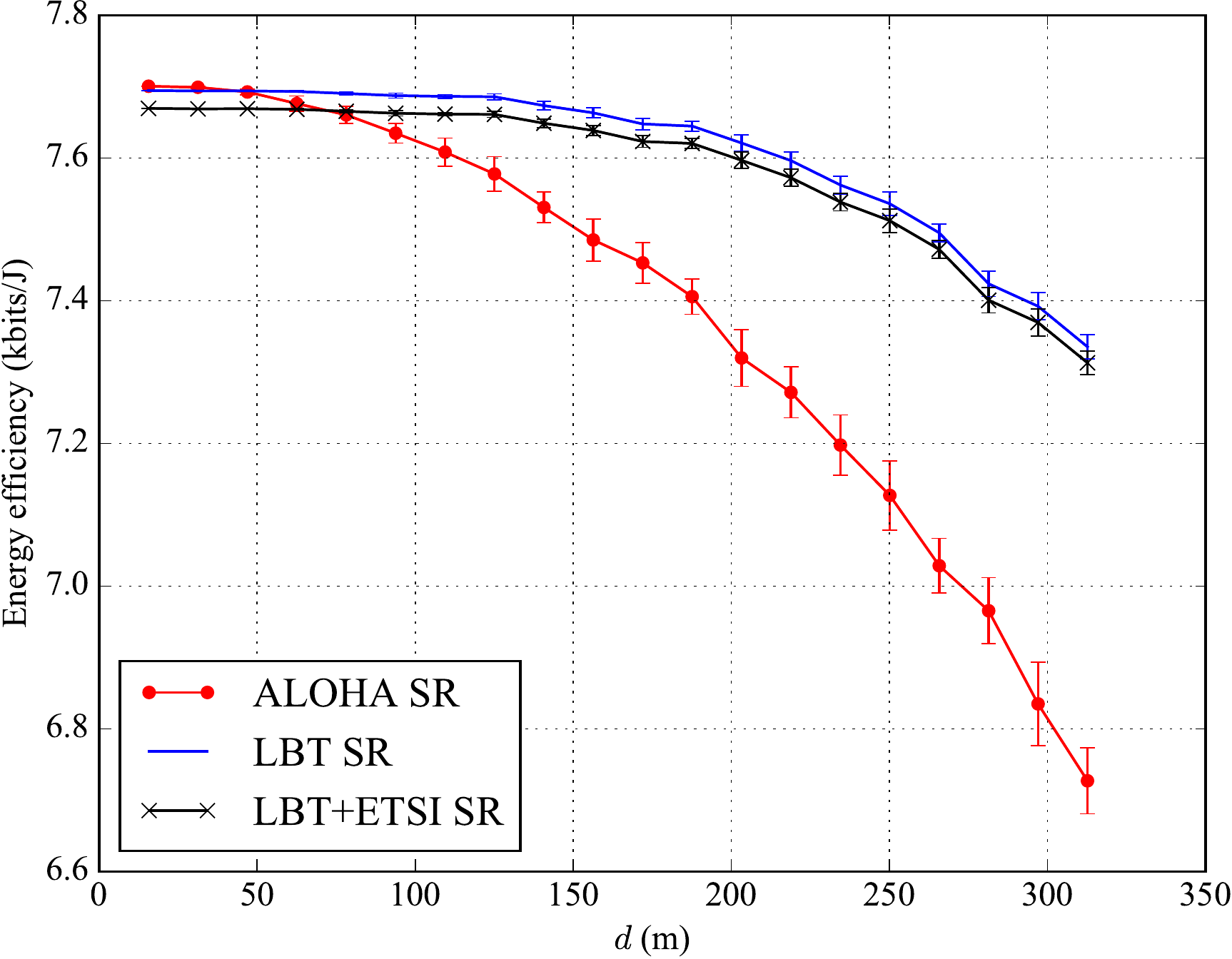}
		\caption{Single rate (SR) case.}
		\label{fig:energy-sr}
	\end{subfigure}\vspace{1em}
	\begin{subfigure}{1\linewidth}
		\centering
		\includegraphics[width=1\linewidth]{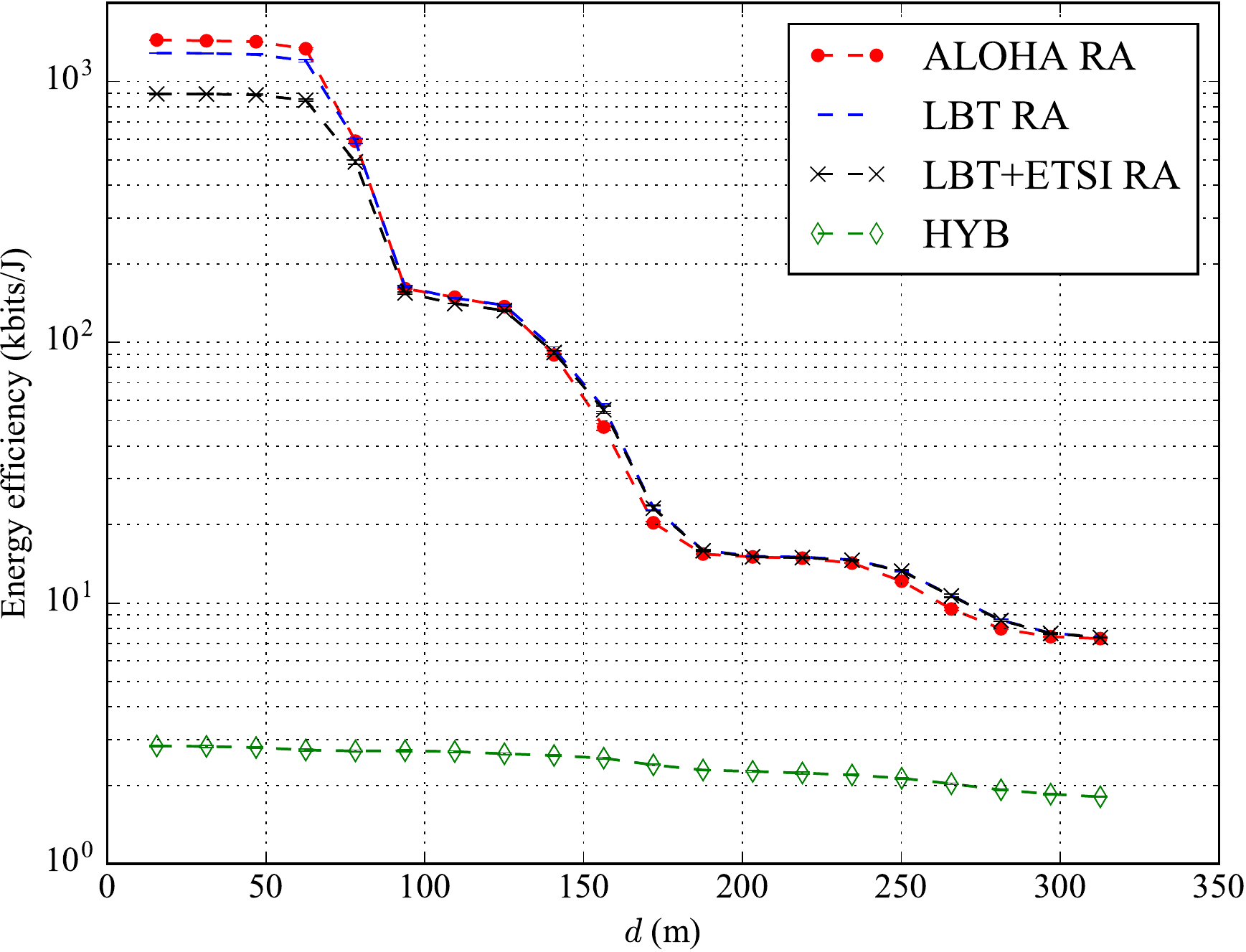}
		\caption{Rate adaptive (RA) case.}
		\label{fig:energy-ra}
	\end{subfigure}
	\label{fig:energy}
	\caption{Successfully received bits per unit of consumed energy, with 95\%~confidence intervals.}
\end{figure}

In  Figure~\ref{fig:energy-sr} we show the energy efficiency for ALOHA and LBT access schemes when varying the distance of the target node from the gateway, in the SR case. We can observe that  peripheral nodes exhibit lower energy efficiency because of the larger number of failure transmissions, and that the carrier sensing mechanism can alleviate this problem. The black curve marked with crosses shows the results obtained when using the parameters imposed by ETSI in the CCA procedure. As it can be seen, the energy efficiency is slightly lower than that obtained with the parameters adopted by commercial technologies, which may suggest that ETSI recommendations in this regard are possibly too conservative.

The adaptive rate case is shown in Figure~\ref{fig:energy-ra}, where we also show the performance achieved by HYB. We can observe that both ALOHA and LBT can reach very high efficiency for nodes near the receiver, since the higher bitrates that decrease the transmit energy and the failure probability. It is worth to note that the first factor is dominant for the energy efficiency. The benefit transfers to the nodes farther away from the gateway, though the performance gain progressively reduces with the distance from the transmitter. 

We also observe that, for nodes closer to the gateway, LBT shows a non-negligible energy efficiency loss with respect to ALOHA, which is even more marked when adopting the ETSI parameters. This is clearly due to the energy cost of the carrier sense mechanism, which takes a time comparable with the packet transmission time when using high bitrates.
Furthermore, as revealed by the analysis of the failure probability, the carrier sense mechanism is not really worth for nodes close to the gateway when using RA,
considering also that it may yield packet drops due to the impossibility of finding the channel idle within the maximum number of carrier sensing attempts. This problem would be further exacerbated in case of overlapping cells. Therefore, the use of CCA appears to be fruitless, if not detrimental, for nodes close to the gateway when RA is enabled. 

Finally, we observe that the energy efficiency of HYB is the worst, being affected by both the higher failure probability observed in Figure~\ref{fig:pfail} and the higher energy consumption due to the transmission of resource messages and the reception of beacons. This inefficiency is more marked for nodes near the receiver, where the energy spent on control messages is actually greater than that used for the high-rate transmissions of small data packets.

\subsection{Coexistence issues}
\begin{figure}
	\centering
	\includegraphics[width=1\linewidth]{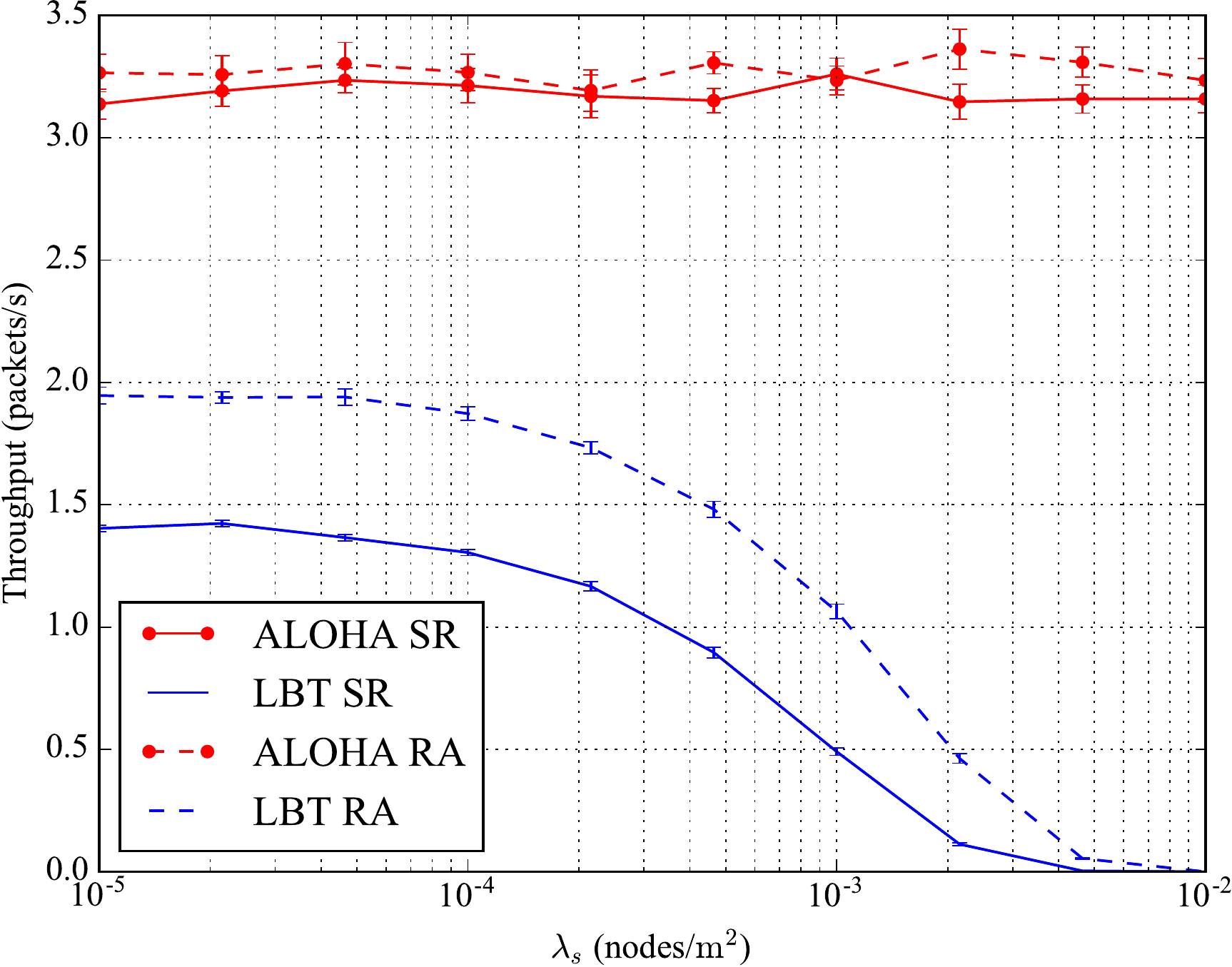}
	\caption{Aggregated throughput for each channel access method in the single and adaptive rate scenarios, with 95\%~confidence intervals.}
	\label{fig:coexistence}
\end{figure}
\begin{figure}
	\centering
	\includegraphics[width=1\linewidth]{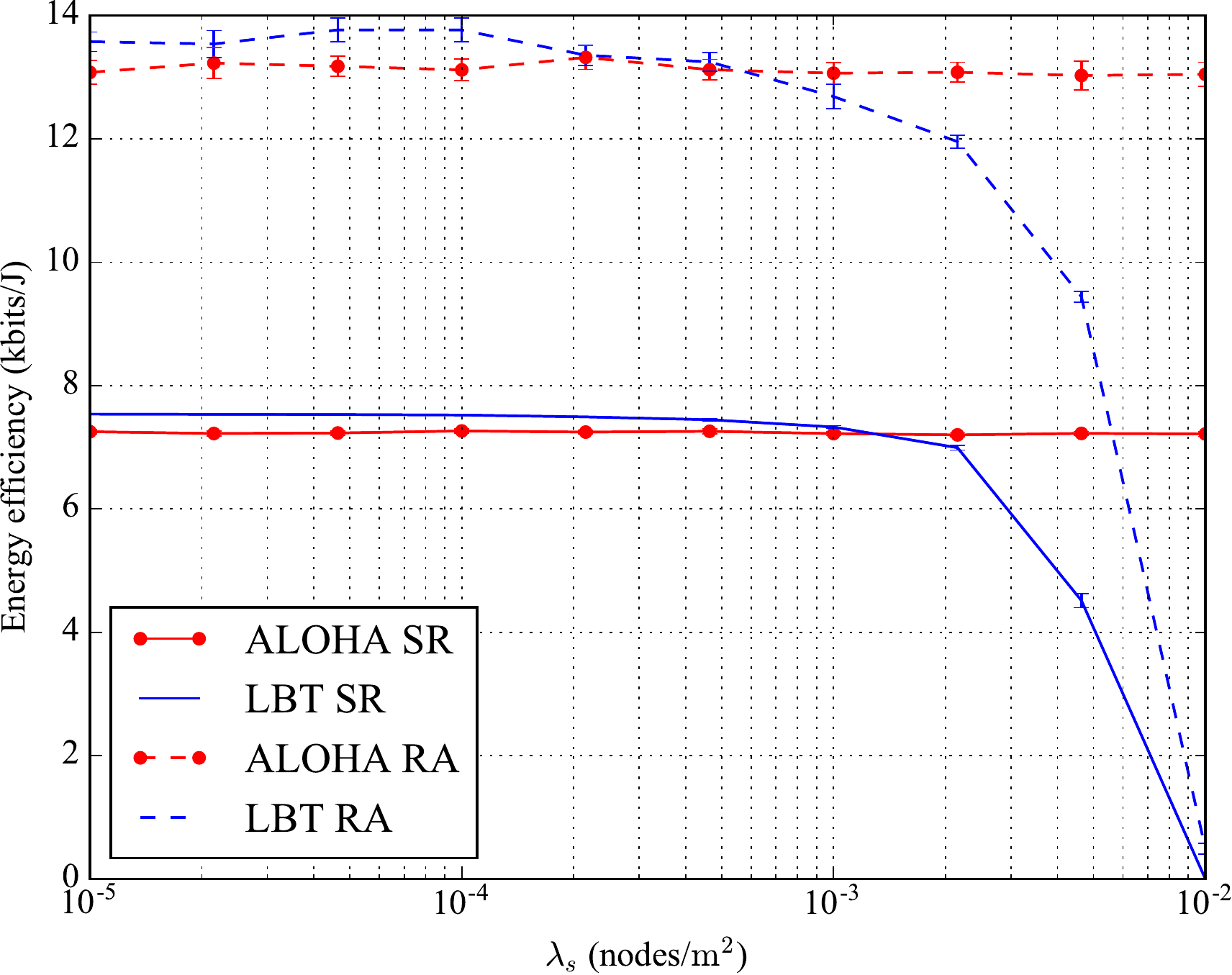}
	\caption{Succesfully received bits per unit of consumed energy for each channel access method, in the single and adaptive rate scenarios, with 95\%~confidence intervals.}
	\label{fig:coexistence-energy}
\end{figure}
Another important question regards the coexistence in the same area of nodes using LBT and ALOHA access schemes.

Figure~\ref{fig:coexistence} and Figure~\ref{fig:coexistence-energy} report the throughput of the two access
methods, defined as the overall rate of successful packet transmissions, and the energy efficiency.
Curves for ALOHA (respectively LBT) have been obtained by fixing the spatial density of this type of nodes to 0.001 nodes/$\mathrm{m}^2$ and increasing the spatial density of LBT (respectively ALOHA) nodes from $10^{-5}$ to $10^{-2}$ nodes/$\mathrm{m}^2$. 

Results in Figure~\ref{fig:coexistence} show that the performance of ALOHA nodes is not impacted by an increase in the number of LBT nodes, while the latter suffer strong performance degradation due to the CCA mechanism that aborts a transmission attempt when the channel is sensed busy for a given number of successive attempts. We can also see that the use of multiple transmission rates can only slightly alleviate the problem, but the fragility of the LBT mechanism in presence of ALOHA traffic still remains. Similar observations can be drawn for the energy efficiency results.
In both cases, the use of RA improves the energy efficiency quite significantly.

\section{Conclusions}\label{con}
In this work, we presented an overview of the three main uncoordinated channel access sensing schemes, namely pure ALOHA, HYB, and LBT, in an IoT scenario.
We compared the performance of these schemes in terms of probability of successful transmission and energy efficiency, by considering the duty-cycle limitation for ALOHA, the control packets for HYB, and the CCA procedure for LBT as mandated by the international regulation frameworks. 

From this analysis, it appears clear that adding rate adaptation capabilities is pivotal to maintain reasonable level of performance when the coverage range and the cell load increase.  Moreover, we observed that LBT generally yields lower transmission failure probability, though packet dropping events may occur because the channel is sensed busy for a certain number of consecutive CCA attempts. This  impacts on the actual energy efficiency of the LBT access scheme, which may turn out to be even smaller than that achieved by ALOHA schemes. Furthermore, we also observed that LBT performance undergoes severe degradation when increasing the number of ALOHA devices in the same cell, again because of the channel-blockage effect caused by the other transmitters. Finally, the HYB scheme proves ineffective in the considered scenario, since the reservation channel becomes the system bottleneck with short data packets. Nonetheless, hybrid solutions that adopt LBT for peripheral nodes and ALOHA for nodes closer to the receiver, or apply rate adaptation also to the reservation phase, can potentially lead to a general performance improvement of the system. This analysis, however, is left to future work.

\bibliographystyle{IEEEtran}
\bibliography{main_submitted}

\begin{IEEEbiographynophoto}{Daniel Zucchetto}
 received the Bachelor degree in
Information Engineering in 2012 and the Master degree
in Telecommunication Engineering in 2014, both from  the
University of Padova, Italy. Since October 2015 he is a Ph.D.
student at the Department of Information Engineering of
the University of Padova, Italy. His research interests include
Low-Power Wide-Area Network technologies and next generation cellular
networks (5G), with particular focus on their application to
the Internet of Things.
\end{IEEEbiographynophoto}

\begin{IEEEbiographynophoto}{Andrea Zanella}
 (S'98-M'01-SM'13) is Associate Professor at the University of Padova, Padova, Italy. He has authored more than 130 papers, four books chapters and three international patents in multiple subjects related to wireless networking and Internet of Things. Moreover, he serves as Editor for many journals, included the IEEE Internet of Things Journal, and the IEEE Transactions on Cognitive Communications and Networking.
\end{IEEEbiographynophoto}

\end{document}